\documentclass[runningheads]{llncs}
\usepackage{graphicx}

\usepackage[ngerman, english]{babel}

\usepackage[utf8]{inputenc}
\usepackage[T1]{fontenc}
\usepackage{times} 
\usepackage{url} 
\usepackage{latexsym} 
\usepackage{tipa} 
\usepackage[autostyle]{csquotes}
\usepackage{relsize}
\usepackage[dvipsnames]{xcolor}
\usepackage[ruled, linesnumbered]{algorithm2e} 
\usepackage{fancyvrb} 
\usepackage{amssymb}
\usepackage{upquote}
\usepackage[autostyle]{csquotes}

\usepackage{mathptmx}
\usepackage{amsmath}
\usepackage{textcomp}

\usepackage{siunitx}
\usepackage{graphicx}
\usepackage{rotating} 
\usepackage{tikz}
\usetikzlibrary{calc}
\usetikzlibrary{matrix}
\usetikzlibrary{positioning}
\usetikzlibrary{shapes.callouts}
\usepackage{adjustbox}

\usepackage{enumitem}
\setlist{nosep}
\setlist[description]{leftmargin=\parindent}

\usepackage{tabularx}
\usepackage{tabulary}
\usepackage{booktabs}
\usepackage{colortbl}
\usepackage{capt-of}
\usepackage{array} 

\usepackage[singlelinecheck=off, justification=raggedright]{subcaption}
\usepackage[hidelinks, pdfusetitle]{hyperref}
\hypersetup{pdfauthor={Eran Raveh, Ingmar Steiner, Iona Gessinger, Bernd Möbius}}

\usepackage[capitalise, nameinlink, noabbrev]{cleveref}
\crefformat{footnote}{#2\footnotemark[#1]#3}

\usepackage[obeyFinal]{todonotes}
\usepackage{acro}
\DeclareAcronym{asp}	{short=ASP,		long=additional speech processing}
\DeclareAcronym{asr}	{short=ASR, 	long=automatic speech recognition, long-plural=s, short-plural=s}
\DeclareAcronym{call}	{short=CALL, 	long=computer-assisted language learning, long-plural=s, short-plural=s}
\DeclareAcronym{capt}	{short=CAPT, 	long=computer-assisted pronunciation training, long-plural=s, short-plural=s}
\DeclareAcronym{cat}	{short=CAT, 	long=communication accommodation theory, long-plural=s, short-plural=s}
\DeclareAcronym{dfg}	{short=DFG,		long=German Research Foundation}
\DeclareAcronym{dm}		{short=DM,		long=dialogue manager, long-plural=s, short-plural=s}
\DeclareAcronym{gui}	{short=GUI, 	long=graphical user interface, long-plural=s, short-plural=s}
\DeclareAcronym{hci}	{short=HCI,		long=human-computer interaction}
\DeclareAcronym{hhi}	{short=HHI, 	long=human-human interaction, long-plural=s, short-plural=s}
\DeclareAcronym{hmm}	{short=HMM, 	long=hidden markov model, long-plural=s, short-plural=s}
\DeclareAcronym{hts}	{short=HTS, 	long=\ac{hmm} based Speech Synthesis System}
\DeclareAcronym{its}	{short=ITS, 	long=intelligent tutoring system, long-plural=s, short-plural=s}
\DeclareAcronym{nlg}	{short=NLG, 	long=natural language generation, long-plural=s, short-plural=s}
\DeclareAcronym{nlp}	{short=NLP, 	long=natural language processing, long-plural=s, short-plural=s}
\DeclareAcronym{nlu}	{short=NLU, 	long=natural language understanding, long-plural=s, short-plural=s}
\DeclareAcronym{rmse}	{short=RMSE, 	long=root-mean-square error}
\DeclareAcronym{sds}	{short=SDS, 	long=spoken dialogue system, long-plural=s, short-plural=s}
\DeclareAcronym{smo}	{short=SMO, 	long=sequential minimization optimization, long-plural=s, short-plural=s}
\DeclareAcronym{svm}	{short=SVM, 	long=support vector machine, long-plural=s, short-plural=s}
\DeclareAcronym{tts}	{short=TTS, 	long=text-to-speech}

\graphicspath{{figures/}}

\begin{document}

	\hyphenation{%
		CMU-Sphinx
	}
	
	\title{Studying Mutual Phonetic Influence\\
		with a Web-Based Spoken Dialogue System
	}
	%
	%
	\author{Eran Raveh\inst{1,2}\orcidID{0000-0003-4411-9663} \and
		Ingmar Steiner\inst{1-3}\orcidID{0000-0001-6415-5915} \and \\
		Iona Gessinger\inst{1,2}\orcidID{0000-0001-5333-9794} \and
		Bernd Möbius\inst{1}\orcidID{0000-0003-3065-9984}}
	\authorrunning{E.\ Raveh et al.}
	%
	\institute{Language Science \& Technology, Saarland University, Saarbrücken, Germany \and
		Multimodal Computing and Interaction, Saarland University, Saarbrücken, Germany \and
		German Research Center for Artificial Intelligence (DFKI GmbH), Saarbrücken, Germany
		\email{raveh@coli.uni-saarland.de}}
	%
	\maketitle              

	\setcounter{footnote}{0}
	
	\begin{abstract}
		This paper presents a study on mutual speech variation influences in a human-computer setting.
The study highlights behavioral patterns in data collected as part of a shadowing experiment,
and is performed using a novel end-to-end platform for studying phonetic variation in dialogue.
It includes a spoken dialogue system capable of detecting and tracking the state of phonetic features in the user's speech and adapting accordingly.
It provides visual and numeric representations of the changes in real time,
offering a high degree of customization, and can be used for simulating or reproducing speech variation scenarios.
The replicated experiment presented in this paper along with the analysis of the relationship between the human and non-human interlocutors lays the groundwork for a spoken dialogue system with personalized speaking style, which we expect will improve the naturalness and efficiency of human-computer interaction.
		\keywords{Spoken dialogue systems \and Phonetic convergence \and Human-computer interfaces.}
	\end{abstract}
	\section{Introduction}
	\label{sec:introduction}
	
	With expanding research on, and growing use of, \acp{sds}, a main challenge in the development of \ac{hci} systems of this kind is making them as close as possible to \ac{hhi} in terms of naturalness, fluency, and efficiency.
	One aspect of such \acp{hhi} is the relationship of mutual influences between the interlocutors.
	Influence here means changes in one interlocutor's conversational behavior triggered by the  behavior of the other interlocutor.
	We refer to changes that make the interlocutors' behaviors more similar as \emph{convergence}.
	Convergence can occur in different modalities and with respect to various aspects of the conversation, like eye gaze, gestures, lexical choices, body language, and more.
	In this paper, we concentrate on phonetic-level influences, i.e., \emph{phonetic convergence}.
	More specifically, we examine pronunciation variations over the course of \acp{hci}.
	As speech is the principal modality used for interacting with \acp{sds}, we believe it is an especially important modality to study in the field of \ac{hci}.
	Simulating and triggering convergence on the phonetic level, as found in \ac{hhi}, may contribute a lot to the naturalness of dialogues of humans with computers.
	\Acp{sds} with such personalized speech style are expected to offer more natural and efficient interactions, and move one more step away from the \emph{interface metaphor} \cite{Edlund/etal:2006} toward the \emph{human metaphor} \cite{Carlson/etal:2006}.
		
	The novel system introduced in \cref{sec:system} tracks the states of segment-level phonetic features during the dialogue.
	All of the analyses are automated and run in real time.
	This not only saves time and manual work typically needed in convergence studies, but also makes the system more suitable for integration into other applications.
	In \cref{sec:showcase}, we use this newly introduced system with recordings collected as part of a shadowing experiment to examine the relationship of mutual influences between a (simulated) user and the system.
	Using these signals, the system provides both visual and numerical evidence of the mutual influences between the interlocutors over the course of the interaction.
    The system itself will be made freely available under an open-source license.

	\section{Background and Related Work}
	\label{sec:background_and_related_work}
	
	Integrating support for changes in the speech signal into computer systems may enhance \ac{hci} and provide improved tools for studying convergence in \ac{hci}.
	\cite{Oviatt/etal:2004} discusses the advantages of systems that dynamically adapt their speech output to that of the user, and the challenges involved in developing and using these systems.
	
	\subsection{Phonetic Convergence}
	\label{subsec:phonetic_convergence}
	
	According to \cite{Pardo:2006}, phonetic convergence is defined as an \emph{increase in segmental} \emph{and suprasegmental} \emph{similarity between two interlocutors} (e.g., \cite{Walker/Campbell-Kibler:2015}).
	In contrast to \emph{entrainment}, we use the term \emph{convergence} to describe dynamic, mutual, and non-imposing changes.
	Phonetic convergence has been found to various extent in conversational settings \cite{Lewandowski:2012}.
	There is evidence for phonetic convergence being both an internal mechanism \cite{Pickering/Garrod:2004} and socially motivated \cite{Kim/etal:2011}.
	Previous studies of phonetic convergence in spontaneous dyadic conversations have focused on speech rate \cite{Schweitzer/Walsh:2016}, timing-related phenomena \cite{Putman/Street:1984}, pitch \cite{Gessinger/etal:2018}, intensity \cite{Levitan/Hirschberg:2011}, and perceived attractiveness \cite{Michalsky/Schoormann:2017}.
	Phonetic convergence is often examined in the scope of shadowing experiments, in which the participants are asked to produce certain utterances after hearing them produced in some stimuli (e.g., \cite{Gessinger/etal:2017}).
	This is typically done with single target words embedded in a carrier sentence.
	The experiment showcasing our system in \cref{sec:showcase} uses whole sentences as stimuli, in which the target features are embedded, making it a semi-conversational \ac{hci} setting.
	
	\subsection{Adaptive Spoken Dialogue Systems}
	\label{subsec:adaptive_sds}
	
	Various studies have investigated entrainment and priming in \acp{sds}, aiming to better understand \ac{hci} dynamics and improve task-completion performance.
	\cite{Lopes/etal:2013}, for example, focused on dynamic entrainment and adaptation on the lexical level.
	Others, like \cite{Nenkova/etal:2008}, concentrated on word frequency.
	\cite{Parent/Eskenazi:2010} examined changes in both lexical choice and word frequency.
	While these studies addressed the changes in experimental, scripted scenarios, the theoretical foundations for studying these changes in spontaneous dialogue exist as well \cite{Brennan:1996}.
	\cite{Gasic/etal:2013} provide examples of online adaptation for dialogue policies and belief tracking.
	
	It is important to note that while all of the studies mentioned above examine various aspects of dialogues, none of those are related to speech -- the primary modality used to interact with \acp{sds}.
	Studying convergence of speech in an \ac{hci} context is made possible with more natural synthesis technology, which gives fine-grained control over parameters of the system's spoken output.
	Many systems that deal with adaptation of speech-related features focus on prosodic characteristics like intonation or speech rate.
	\cite{Levitan:2014} sheds light on acoustic-prosodic entrainment in both \ac{hhi} and \ac{hci} via the use of interactive avatars.
	\cite{Bell/etal:2003} found that users' speech rate can be manipulated using a simulated \ac{sds}.
	Similar results were found when intensity changes in children's interaction with synthesized \ac{tts} output were examined \cite{Coulston/etal:2002}.
	
	All of the above provide solid ground for further investigation of phonetic convergence in \ac{hci} using \acp{sds}.
	
	\section{System}
	\label{sec:system}
	
	\begin{figure*}[t]
		\centering
		\includegraphics[width=0.9\linewidth]{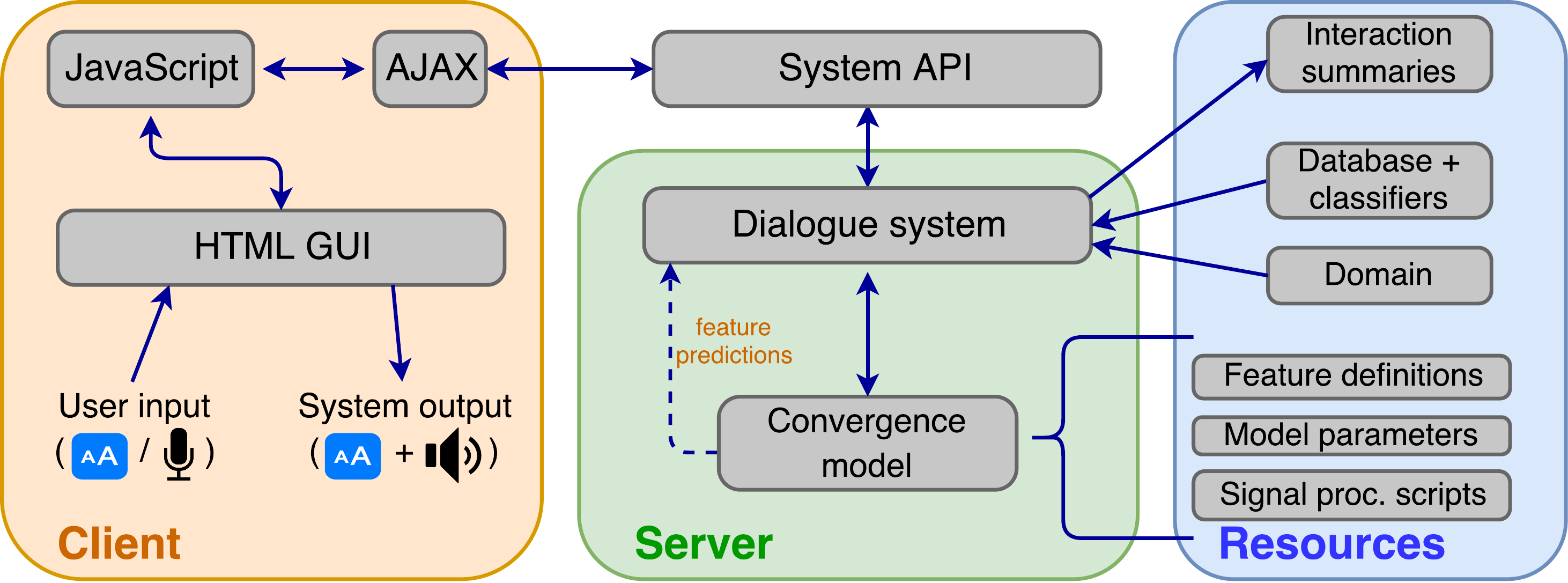}
		\caption{An overview of the system architecture.
			The background colors distinguish client components, server components, and external resources that can be customized.}
		\label{fig:architecture}
	\end{figure*}
	
	The system introduced here is an end-to-end, web-based \ac{sds} with a focus on phonetic convergence and its analysis over the course of the interaction.
	Besides placing convergence in the spotlight, it is designed to be flexible and to meet the researcher's needs by offering a wide range of customizations (see \cref{subsec:models_and_cusomizations}).
	Its online access via a web browser makes it scalable and simple for the end-user to operate.
	The system's architecture and functionality are described in \cref{subsec:architecture},
	its \ac{gui} and operation in \cref{subsec:graphical_user_interface},
	and an example of its utilization is demonstrated in \cref{sec:showcase}.
	Ultimately, it offers an experimentation platform for studying phonetic convergence, with emphasis on the following:
	
	\begin{description}
		\item[Temporal~analysis]
		offering real-time visualization of the interlocutors' relations with respect to selected phonetic features over the course of the interaction.
		
		\item[Customizability]
		allowing the user to experiment with different scenarios by configuring parameters and definitions in many of the system's components.
		
		\item[Online~scalability]
		connecting multiple web clients to a server, allowing users to use it anywhere without preceding installation and configurations, and helping experimenters to collect and replay acquired data.
	\end{description}
	
	\subsection{Architecture}
	\label{subsec:architecture}
	
	As the system aims to offer a customizable playground for experimenting and studying phonetic convergence in \ac{hci}, a key aspect of its architecture is the separation between client-side, server-side, and external resources (see \cref{fig:architecture}).
	All of the resources and configuration files needed for designing the interaction are located on the server.
	Running the client and server on different machines allows users to interact with the system using a web browser alone.
	
	\begin{figure}[t]
		\centering
		\includegraphics[width=0.9\linewidth]{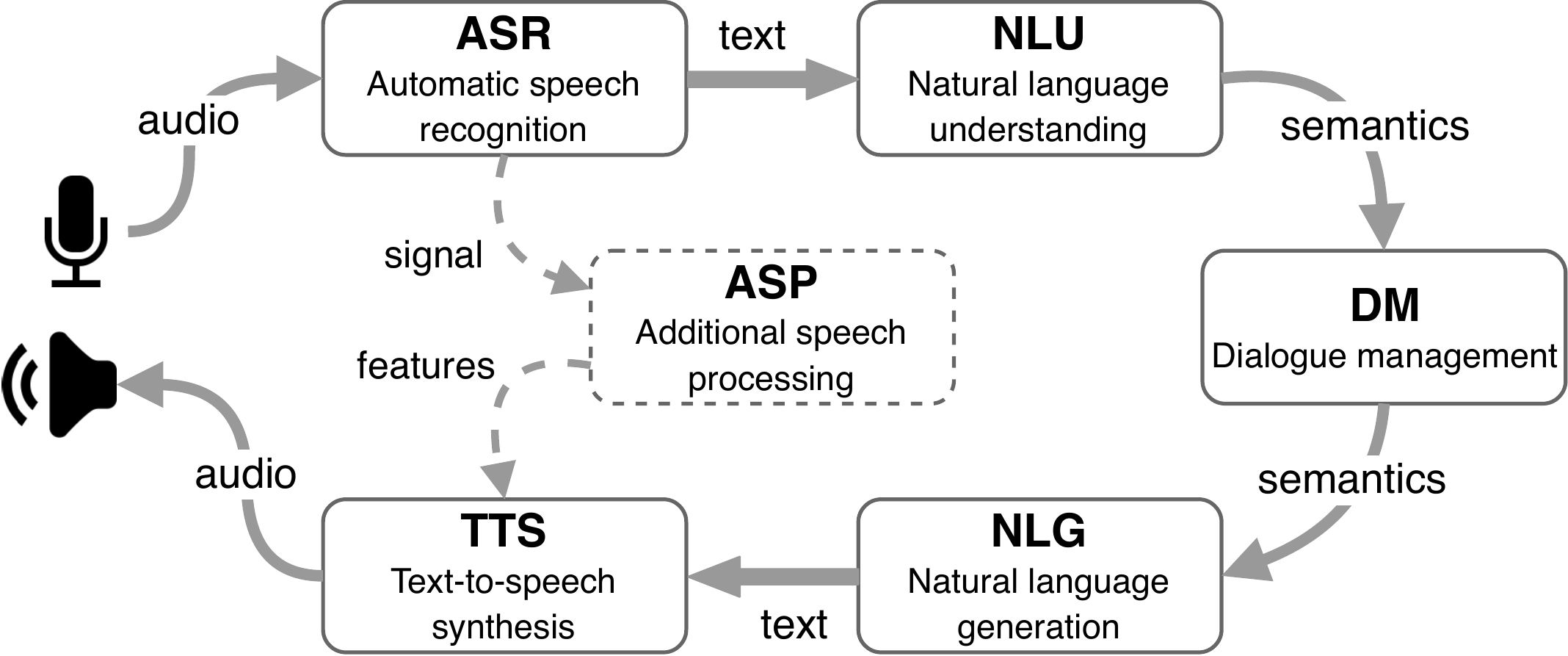}
		\caption{The architecture of the dialogue system component.
			The \acs{asp} module (dashed line) between the \ac{asr} and \ac{tts} modules is responsible for performing additional speech processing required for analyzing the phonetic changes.
			Though additional links between the \ac{asp} module and other modules (like \ac{nlg} for example) could be made, those are beyond the scope of this work.}
		\label{fig:architecture_sds}
	\end{figure}
	
	As shown in \cref{fig:architecture_sds}, the \textbf{dialogue system} component consists of typical \ac{sds} modules such as \ac{nlu} and a \ac{dm}, but also contains an \acf{asp} module \cite{Raveh/Steiner:2017c}.
	This module is responsible for processing the audio and extracts the features required by the convergence model.
	While the \ac{nlu} component uses merely the transcription provided by the \ac{asr}, the \ac{asp} module analyzes the speech signal itself.
	More specifically, it tracks occurrences of the defined features and passes their measured values to the convergence model, which, in turn, forwards the tracked feature parameters to the \ac{tts} synthesis component.
	
	\subsection{Models and Customizations}
	\label{subsec:models_and_cusomizations}
	
	The \textbf{computational model} for phonetic convergence used in the system is described in \cite{Raveh/etal:2017b}.
	Different phonetic convergence behavioral patterns that were observed in \ac{hhi} and \ac{hci} experiments can be simulated by combinations of the model's parameters presented in \cref{tab:convergence_model}.
	All of the parameters can be modified in the system's configuration file.
	
	\begin{table}[t]
		\centering
		\caption{Summary of the computational model's parameters in their order of application in the convergence pipeline.
			Parameters marked with an asterisk \enquote*{*} are defined for each feature independently.}
		\vspace{-.8cm}
		\noindent\rule[-1cm]{\linewidth}{1pt} 
		\begin{description}[labelwidth=\widthof{\quad \bfseries calculation method*}, style=multiline, leftmargin=\labelwidth]
			\item[allowed range*] allowed value range for new instances
			\item[history size] maximum number of exemplars in pool
			\item[update frequency] frequency to recalculate feature's value
			\item[calculation method*] method to calculate pool value
			\item[convergence rate] weight given to pool value when recalculating
			\item[convergence limit*] the maximum degree of convergence allowed
		\end{description}
		\noindent\rule[1cm]{\linewidth}{1pt} 
		\vspace{-1.6cm}
		\label{tab:convergence_model}
	\end{table}
	
	The entire convergence process is based on the the \textbf{tracked phonetic features} that are considered \enquote{convergeable}, i.e., prone to variation, and is triggered whenever the \ac{asr} component detects a segment containing a phoneme associated with one or more of these features.
	Each feature is defined by a key-value map, in which the parameters from \cref{tab:convergence_model} are configured.
	A classifier can be associated with each feature to provide real-time predictions for both the user's and the system's realizations of that feature, as demonstrated in \cref{fig:plot}.
	With this information available, more meaningful insights can be gained into the dynamics of phonetic changes in the dialogue.
	
	The \textbf{dialogue domain} is specified in an XML-based file.
	More details on the domain file can be found in \cite{Lison/Kennington:2015}.
	The format of the domain file makes it easy to define new scenarios for the system,
	such as a task-specific dialogue, general-purpose chat, or an experimental setup.
	
	\textbf{Speech processing} is a central aspect of the system.
	Different models can be used, e.g., for improving performance or changing the language or the \ac{asr} module or the output voice of the \ac{tts} module.
	
	\subsection{Graphical User Interface}
	\label{subsec:graphical_user_interface}
	
	The system's \ac{gui} consists of three main areas:
	\begin{figure}[t]
		\centering
		\includegraphics[width=0.7\linewidth]{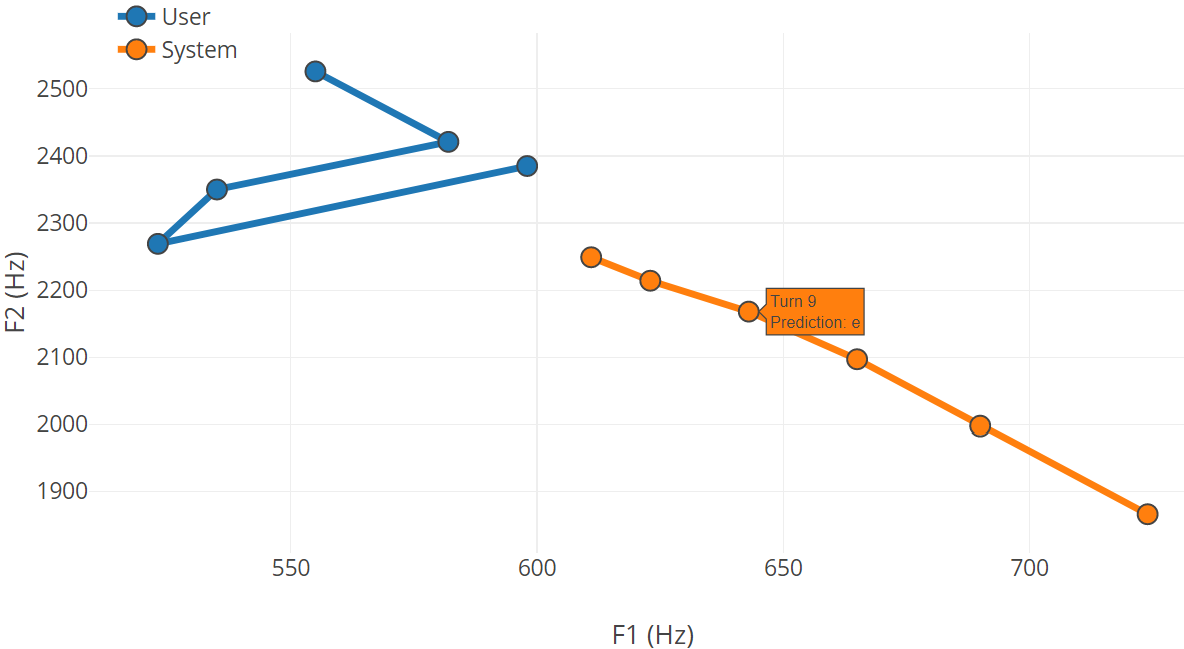}
		\caption{A screenshot of the plot area showing the states of the feature \textipa{[E:]}~vs.~\textipa{[e:]} (in 		2-dimensional formant space) during an interaction.
			The system's internal convergence model (orange, bottom right) gradually adapts to the user's (blue, upper left) detected realizations.
			A prediction of the feature's current realization is given for both interlocutors.
			The annotation box marks the turn in which the system has aggregated enough evidence from the user's utterances and changes its pronunciation from \textipa{[E:]} (its initial state) to \textipa{[e:]} (the user's preferred variation).}
		\label{fig:plot}
	\end{figure}
	
	In the \textbf{chat area}, the interaction between the user and the system is shown in a chat-like representation.
	Each turn's utterance appears inside a chat bubble with different colors and orientations for the user and the system.
	The turns are also numbered, to better track the dialogue progress and analysis shown by the plots in the graph area.
	It is also possible to replay the utterance of a turn by clicking the \enquote{Play} button in its corresponding bubble.
	
	In the \textbf{interaction area}, the user can interact with the system with written or spoken input.
	Text-based interactions progress through the dialogue (if applicable) and trigger any subsequent domain model, but will not affect convergence-related models, since there is no audio input to process.
	Spoken input can be provided either by speaking into the microphone or via audio files with pre-recorded speech.
	The latter option is especially useful for simulating specific user input, or for reproducing a previous experiment, as done in \cref{sec:showcase}.
	
	In the \textbf{graph area}, each of the tracked features is visualized in a separate plot, and new data points are added whenever a new instance of the feature is detected.
	Hovering over a data point in a graph reveals additional information, such as the turn in which it was added, or the realized variant of the feature produced in that turn as predicted by its classifier.
	These dynamic, interactive plots make it possible to shed light on how the interlocutors influence each other, whether or not they are aware of it, throughout their exchanges.
	\Cref{fig:plot} shows such a graph with several accumulated data points.
	
	\section{Showcase: Examining Convergence Behaviors}
	\label{sec:showcase}
	
	For demonstrating a possible use of the system, we simulated the shadowing experiment detailed in \cite{Gessinger/etal:2017} using the system and its analyses to look into types of participant convergence behavior with respect to the features examined in the experiment (see \cref{tab:target_features}).
	This experiment is designed to trigger phonetic convergence by confronting the participants with stimuli in which certain phonetic features are realized in a manner different from their own realizations.
	The simulation was carried out by building a domain file with the experimental procedure, including the transition between the experiment's phases, as well as the flow within each phase.
	This automates the procedure and adapts it to the participant's pace.
	Participants were simulated by using their recorded speech from the original experiment in the same order.
	The use of the system for this purpose results in an automated, reproducible execution, with additional insights like classification of feature realizations and dynamic visualizations in the \ac{gui}.
	The classifiers were trained offline on the data points acquired from analyzing the stimuli.
	However, the system also supports incremental, online re-training whenever requested by the user,
	for example after every time the convergence model is updated.
	For the demonstration presented here, a \ac{smo} \cite{Platt:1999} implementation of the \ac{svm} classifier was used for training.
	Each turn's number and prediction are added as an interactive annotation to the dynamic graph of the relevant features, as shown in \cref{fig:plot}.
	Finally, using the system, the experiment is transformed into an automated dialogue scenario, which enhances its \ac{hci} nature.
	
	\begin{table}[t]
		\centering
		\caption{Examples of stimuli sentences, each containing one target feature.}
		\begin{tabularx}{\linewidth}{@{}*{5}{l}@{\qquad\qquad	}l}			
			\multicolumn{5}{c}{\bfseries{Sentence}} & \multicolumn{1}{c}{\textbf{Feature}} \\
			
			\toprule
			
			War          & das          & Ger\textbf{\underline{ä}}t & sehr          & teuer? & \textipa{[E:]}~vs.~\textipa{[e:]} in word-medial $\langle$ä$\rangle$ \\
			\emph{Was} & \emph{the} & \emph{device}            & \emph{very} & \emph{expensive?} & \\[0.1cm]
			
			Ich          & bin         & sücht\textbf{\underline{ig}} & nach        & Schokolade. & \textipa{[I\c{c}]}~vs.~\textipa{[Ik]} in word-final $\langle$-ig$\rangle$ \\
			\emph{I}   & \emph{am} & \emph{addicted}            & \emph{to} & \emph{chocolate.} & \\[0.1cm]
			
			Wir         & besuch\textbf{\underline{en}} & euch         & bald          & wieder. & \textipa{[\s{n}]}~vs.~\textipa{[@n]} in word-final $\langle$-en$\rangle$ \\
			\emph{We} & \emph{will visit}           & \emph{you} & \emph{soon} & \emph{again.} & \\
			
			\bottomrule
		\end{tabularx}
		\label{tab:target_features}
	\end{table}
	
	\subsection{Finding Behavioral Patterns}
	\label{subsec:validation}
	
	In this section, we focus on the validation for the feature \textipa{[E:]}~vs.~\textipa{[e:]} as a representative example for the phonetic adaptation capability of the system.
	Although the classified realization is binary (\textipa{[E:]}~or~\textipa{[e:]}), the underlying representation used by the model is gradual.
	Both of these views on the feature can be seen in the graph area, as shown in \cref{fig:plot}. 
		
	The degree of convergence was examined per utterance in the shadowing phase of the experiment.
	Three main groups emerged, each with a different behavior:
	one group of participants showing little to no tendency to converge (changes in \SI{\le 10}{\percent} of their utterances),
	the second, with varying degrees of convergence (\SIrange{10}{90}{\percent}),
	and a third group of participants who were very sensitive to the stimuli's variation (\SI{\ge 90}{\percent}).
	We refer to these groups as \emph{Low}, \emph{Mid}, and \emph{High}, respectively.
	The feature's classifier was determined on the fly, so that the prediction for each utterance was made based on the type of the stimulus to which the participant was listening.
	As \cref{tab:validation_shadow} shows, the \emph{Low} and \emph{High} groups are both of significant size, indicating that these two distinct behaviors exist in the data and can be spotted by the system.
	
	In addition, we validated the separation between these behaviors.
	To this end, we regarded the shadowing phase as an annotation task, where the annotators are the predictors of the user and the system.
	Note that \SI{100}{\percent} similarity would mean complete convergence to every stimulus, which cannot be reasonably expected (cf.\ \cite{Gessinger/etal:2017}).
	The Cohen's kappa ($\kappa$) values%
	\footnote{as calculated by the \emph{kappa2} command of the \emph{irr} R package (v0.84), \url{https://cran.r-project.org/package=irr}}
	of the \emph{Low} group are expected to be the lowest, as a lesser degree of convergence was found among these participants.
	By the same logic, the \emph{High} group is expected to have the highest agreement, and the \emph{Mid} to have values between the two other groups.
	Indeed, this hypothesis holds:
	weak agreement was found in the \emph{Low} group, strong agreement in the \emph{High} group, and a value close to \num{0} (indicating no consistent behavior) for the \emph{Mid} group.
		
	\section{Conclusion and Future Work}
	\label{sec:conclusion}
	
	We have introduced a system with an integrated \acf{sds}, which can track and analyze mutual influence on the phonetic level during the interaction based on an internal convergence model.
	This combines work done in the fields of phonetic convergence and adaptive \acp{sds}, and contributes to the understanding of power relations between a human and a computer interlocutors.
	Many aspects of the system are customizable, which makes it flexible in terms of possible supported scenarios.
	The system can also run on a separate server, which makes it easier to scale its online use.
	
	To showcase its capabilities, we simulated a replication of a shadowing experiment, which examined phonetic convergence regarding certain segment-level phonetic features.
	Three main user behaviors were found with respect to their tendency to change their pronunciation based on the system's stimulus input.
	This sheds light on possible relations and dynamics between a user and a system in \ac{hci}.
	Running the experiment in this way not only saved time by automating the annotation and phonetic analysis, but also offered additional insight such as visualization and on-the-fly classification.
	We believe that this shows that phonetic convergence can be studied using our \ac{sds}, and that this is one step forward toward personalized, phonetically aware \acp{sds}, which will enable more natural and efficient interaction.
	
	\begin{table}[t]
		\centering
		\caption{A summary of the measures for similarity and agreement between the predictor annotations of user and model productions in the shadowing phase.}
			\begin{tabular}{lSS@{}lS}
				\toprule
							& \multicolumn{1}{c}{Similarity (\si{\percent})} & \multicolumn{2}{c}{Agreement ($\kappa$)} & \multicolumn{1}{c}{Size (\si{\percent})} \\
				\midrule
				\emph{Low}  & {<}1 & -0.57 & {***} & 23 \\
				\emph{Mid}  &   22 & -0.15 & {*}   & 50 \\
				\emph{High} &   26 &  0.81 & {***} & 27 \\
				All         &   48 & -0.11 & {*}   & 100 \\
				\bottomrule
			\end{tabular}
		\label{tab:validation_shadow}
	\end{table}
	
	Future work will pursue two independent directions.
	Regarding phonetic convergence, supporting more features will make the system more comprehensive and useful for studying a wider range of phenomena.
	Specifically, adding support for supra-segmental features will enable replication of experiments similar to e.g., \cite{Levitan/etal:2016} in the same manner as in \cref{sec:showcase}.
	As for user acceptance, it would be interesting to examine whether users show any preference toward an \ac{sds} that converges to their speech on the phonetic level, and whether they would change their speaking style based on the system's output, forming an interaction with mutual and dynamic convergence similar to \ac{hhi}.
	The first research question can be tested by comparing user interaction with a baseline system and one with convergence capabilities, and evaluating the users' performance and satisfaction.
	The second research question can be investigated by comparing the users' speech when interacting with either system configuration.
	Additionally, to test the system's influence on users' speech, the users can train with an intelligent \acf{call}, such as a \acf{capt} system, which will change its learner model based on their input.
	Metrics such as task completion rate, performance accuracy, and completion time can be used to evaluate how helpful the system is.
	
	\section*{Acknowledgments}
	
	Funded by the \ac{dfg} under grants STE~2363/1 and MO~597/6.
	
\bibliographystyle{splncs04}
\bibliography{web}

\end{document}